\documentstyle[amssymb,prb,aps]{revtex}

\begin{document}
\draft
\title{Superconductivity in pure and electron doped $MgB_2$: Transport properties
and pressure effects}
\author{B. Lorenz, Y. Y. Xue, R. L. Meng, and C. W. Chu$^*$}
\address{Texas Center for Superconductivity and Department of Physics, \\
University of Houston,\\
Houston, TX 77204-5932, USA}
\address{$^*$also at Lawrence Berkeley National Laboratory, 1 Cyclotron Road,\\
Berkeley, CA 94720, USA\\
and Hong Kong University of Science and Technology, Hong Kong}
\date{\today}
\maketitle

\begin{abstract}
The normal state and superconducting properties of $MgB_2$ and $%
Mg_{1-x}Al_xB_2$ are discussed based on structural, transport, and
high-pressure experiments. The positive Seebeck coefficient and its linear
temperature dependence for $T_c<T<160$ K provide evidence that the
low-temperature transport in MgB$_2$ is due to hole-like metallic carriers.
Structural and transport data show the important role of defects as
indicated by the correlation of T$_c$, the residual resistance ratio, and
the microstrain extracted from x-ray spectra. The decrease of T$_c$ with
hydrostatic pressure is well explained by the strong-coupling BCS theory.
The large scatter of the pressure coefficients of T$_c$ for different MgB$_2$
samples, however, cannot be explained within this theory. We speculate that
pressure may increase the defect density, particularly in samples with large
initial defect concentration.
\end{abstract}






\subsection{Introduction}

The discovery of superconductivity at 40 K in $MgB_{2}$ by Nagamatsu et al. 
\cite{1} has initiated a tremendous amount of experimental and theoretical
work (for a review see Bueza et al.\cite{2}). $MgB_{2}$ appears to be
electrically three dimensional and its grain boundaries have a far less
detrimental effect on the superconducting current. The low magnetic
anisotropy, the large coherence length, and the high critical current
densities in magnetic fields (as compared e.g. to the cuprate
superconductors) recommend this compound for device-related applications.
The low costs of producing bulk ceramics, thin films, tapes, and wires
further reveal the huge potential this compound may have in the near future. 
$MgB_{2}$ crystallizes in the hexagonal $AlB_{2}$ structure. The
honeycomb-like boron layers are separated by the magnesium ions. In analogy
to the graphite lattice $MgB_{2}$ may be considered as a completely
intercalated graphite structure. Early band structure calculations\cite{3}
have shown that Mg substantially donates charges into the boron layers and
that the electronic bands at the Fermi energy derive mainly from the boron $p
$-bands. Thereby, the Fermi surface displays sheets of hole-like as well as
electron-like character. The electrical transport properties in the normal
state are sufficiently well described by the Bloch-Gr\"{u}neisen law of
normal metals. The temperature dependence of the resistivity follows a power
law with exponents between 2 and 3\cite{4,5} indicating that the temperature
dependent scattering process is the electron-phonon scattering. For a simple
binary compound $T_{c}$ appears to be unusually large. Therefore, the
mechanism of superconductivity has been a matter of discussion and different
models have been proposed\cite{3,6}. Within the BCS model the high $T_{c}$
may be explained by the relatively large energy of the boron phonon modes
due to the low mass of the boron ions and a strong electron-phonon
interaction (strong coupling limit, coupling constant $\lambda $). The
phonon-mediated mechanism is supported by a number of experimental results,
notably the observed isotope effects\cite{7}, tunnelling measurements of the
superconducting gap,\cite{8,9,10} and estimates of the electron-phonon
coupling constant ($\lambda \thickapprox 0.7...0.9$) from specific heat\cite
{11} and phonon density-of-states measurements.\cite{12} Calculations of the
electronic and phononic structures \cite{3,13,14,15,16} have shown that the
bond stretching modes of the boron couple strongly to the $p_{x,y}$
electronic bands and the electron-phonon coupling parameter was estimated
between 0.7 and 0.9, lending further support to the strong coupling BCS
theory. In contrast to the BCS theory, Hirsch\cite{6} has proposed an
alternative model where superconductivity in $MgB_{2}$ is driven by the
pairing of dressed holes. In fact, indications for hole-type conduction in
the normal state originally found in the positive thermoelectric power\cite
{17} where confirmed by Hall effect measurements and analogies to high-$T_{c}
$ cuprate superconductors have been discussed. \cite{18} However, one of the
main predictions of this model, an increase of $T_{c}$ with pressure, could
not be confirmed by early high-pressure experiments.\cite{17,19} Soon after
the discovery of superconductivity in pure $MgB_{2}$ attempts have been made
to raise $T_{c}$ by substituting the elements Mg or B by ions of different
size and/or valency (e.g. Al, Li, C, Cu, Be, Na, Zn, Mn, Si, Co, Ni, Fe). No
increase of $T_{c}$ has been reported so far. However, substitutions are a
tool to modify certain microscopic parameters such as lattice constants and
carrier densities and may facilitate the understanding of the microscopic
mechanisms in the normal and superconducting states. Based on the theory of
normal metals we discuss the transport properties, especially the
thermoelectric effect, of $MgB_{2}$ and $Mg_{1-x}Al_{x}B_{2}$ (Section B).
In Section C the influence of defects on $T_{c}$, lattice constants, and
electronic transport is investigated. Pressure effects on the
superconducting transition temperature are discussed in the final Section.

\subsection{Thermoelectric Transport Properties of $MgB_{2}$ and $%
Mg_{1-x}Al_{x}B_{2}$}

Electrical transport measurements of $MgB_{2}$ have been reported in a large
number of papers. The resistivity roughly follows a power law,

\begin{equation}
\rho (T)=\rho _{0}+AT^{n}  \eqnum{1}
\end{equation}

with an exponent $n$ between 2 and 3. Equation (1) actually can be derived
from the generalized Bloch-Gr\"{u}neisen law for temperatures well below the
Debye temperature.\cite{5} The first term in (1), $\rho _{0}$, reflects the
temperature independent scattering mechanisms (e.g. impurity or defect
scattering) and the second term represents the contribution due to phonon
scattering. Below about $100\ K$ the phonon term is already very small as
compared to $\rho _{0}$ indicating that the material is close to the
residual resistance limit. Resistivity data, however, are controversial
because they are strongly affected by scattering at grain boundaries,
defects, and impurities. A measure of these effects may be found in the
residual resistance ratio, $RRR=\rho (300K)/\rho _{0}$, where $\rho _{0}$ in 
$MgB_{2}$ is well approximated by $\rho (40K)$. A small $RRR$ indicates a
high density of defects and impurities. For $MgB_{2}$ the majority of the
published data show $RRR$-values between 1 and 3, but larger values up to 25
have also been reported. The implications of a high defect concentration on
structural, transport, and superconducting properties will be discussed in
Section C. The Seebeck coefficient is far less affected by grain boundaries
and impurities. Furthermore, it reflects the nature of the charge carriers
at the Fermi surface more sensitively than the conductivity. The
thermoelectric power, $S(T)$, of $Mg_{1-x}Al_{x}B_{2}$ is shown in Fig. 1.
The inset clearly shows the decrease of $T_{c}$ with Al-substitution in
accordance with susceptibility\cite{20} as well as resistivity data\cite{21}%
. The most remarkable features of $S(T)$ are the positive sign over the
whole temperature range and the linear temperature dependence from $T_{c}$
to about 160 $K$ ($MgB_{2}$). At higher temperatures $S(T)$ deviates from
the straight line and tends to saturate at room temperature. The sign and
the linear dependence $S\thicksim T$ are interpreted as the diffusion
thermopower of a hole-type normal metal. The hole character of the charge
carriers in $MgB_{2}$ was also seen in the positive Hall number.\cite{18}
Band structure calculations\cite{3} have shown that the Fermi surface
consists of several hole as well as electron like sheets. At low
temperatures ($T<160\ K$) the hole carriers obviously dominate the charge
transport, but at higher $T$ electron-like states add a negative
contribution to $S(T)$. This may explain the deviations from the linearity
above 160 K.\cite{21,22} Neglecting higher order corrections the diffusion
thermopower, $S_{d}$, is described by the Mott formula,\cite{23}

\begin{equation}
S_{d}=\frac{\pi ^{2}k^{2}T}{3e}\left[ \frac{\partial \ln \sigma (\varepsilon
)}{\partial \varepsilon }\right] _{E_{F}}  \eqnum{2}
\end{equation}

$k$ denotes Boltzmann's constant, $e$ the charge of the carriers, $E_{F}$
the Fermi energy, and $\sigma (\varepsilon )$ is a conductivity-like
function for carriers of energy $\varepsilon $. In the residual resistance
region the logarithmic derivative is simply $1/E_{F}$,

\begin{equation}
S_{d}=\frac{\pi ^{2}k^{2}T}{3eE_{F}}  \eqnum{3}
\end{equation}

For hole carriers the Fermi energy, $E_{F}$, refers to the top of the
conduction band. From equation 3 and the measured slope of $S(T)$ ($%
dS/dT=0.042\ \mu V/K^{2}$ for $x=0$) a rough estimate of $E_{F}$ can be
obtained, $E_{F}\thickapprox 0.57\ eV$. This value may be underestimated
since we have not considered yet any additional contribution (e.g. phonon
drag) to $S(T)$. The phonon drag effect results in a non-linear $S\thicksim
T^{3}$ at low temperatures, but this is not seen in the data of Fig. 1. For
metals with high defect or impurity concentration it is expected that phonon
drag is diminished due to the suppression of the phonon heat current.\cite
{23} Furthermore, the accessible temperatures ($T>40\ K$) for the normal
state thermoelectric power appear to be too high to reveal the true low
temperature behaviour of $S(T)$. The extrapolation of the linear $S(T)$ to
zero temperature yields a small negative value, a possible indication of a
''hidden'' phonon drag contribution. Therefore, the slope of $S(T)$ and the
estimated $E_{F}$ should be considered as an upper bound for the diffusion
part and as a lower limit for the Fermi energy, respectively. Assuming that
charge transport below $160\ K$ is mainly due to hole carriers in the $%
\sigma $ bands our estimate for $E_{F}$ is in fair agreement with the
calculated difference between the Fermi energy and the top of the $\sigma $
bands, $0.9\ eV$ .\cite{24} It is more interesting to investigate the
effects of electron doping (Al substitution) on $S(T)$ and $T_{c}$.
According to the calculated band structure\cite{3} (within a rigid band
model) the addition of electrons reduces the Fermi energy (for $\sigma $
holes) and the density of states $N(E_{F})$. Suzuki et al.\cite{24}
calculated a decrease of $E_{F}$ by $17\ \%$ for $Mg_{0.9}Al_{0.1}B_{2}$.
From the data of Fig. 1 we estimate an increase of slope of $S(T)$ in the
linear region from $0.042\ \mu V/K^{2}$ ($x=0$) to $0.047\ \mu V/K^{2}$ ($%
x=0.05$) and $0.050\ \mu V/K^{2}$ ($x=0.1$). The corresponding decrease of $%
E_{F}$ by $16\ \%$ ($x=0.1$) is in excellent agreement with the calculated
value. The decrease of the superconducting transition temperature with Al
substitution is explained as a density-of-states effect within the strong
coupling BCS theory. $T_{c}$ is given by the McMillan formula,

\begin{equation}
T_{c}=\frac{\omega _{\ln }}{1.2}\exp \left[ -\frac{1.04\left( 1+\lambda
\right) }{\lambda -\mu ^{\ast }(1+0.62\lambda )}\right]  \eqnum{4}
\end{equation}

$\omega _{\ln }$ is the logarithmically averaged phonon frequency, $\mu
^{\ast }$ the screened Coulomb potential and $\lambda =N(E_{F})\left\langle
I^{2}\right\rangle /M\left\langle \omega ^{2}\right\rangle $ is the
electron-phonon coupling constant. $\left\langle I^{2}\right\rangle $ and $%
\left\langle \omega ^{2}\right\rangle $ are the average square of the
electronic matrix element and the phonon frequency, respectively, and $M$ is
the atomic mass. Band structure calculations have shown that $E_{F}$ of $%
MgB_{2}$ is right at an edge of rapidly decreasing density of states.\cite{3}
The increase of the electron number (or decrease of the hole number) leads
to a decrease of $N(E_{F})$ and $\lambda $ and, according to (4), to the
observed reduction of $T_{c}$.

\subsection{The Influence of Defects on Electronic Transport and $T_{c}$}

Electrical resistivity data for $MgB_{2}$ samples reported by various groups
are qualitatively well described by (1). However, huge differences in the
residual resistance ratio have been reported. Whereas the $RRR$ varies
generally between 2 and 3 values as small as 1 and as large as 25 are
possible. Low $RRR$ values are an indication of a large concentration of
defects and impurities and, within the BCS theory, a decrease of $T_{c}$ is
expected with the increase of defect density.\cite{25}

We have synthesized a large number of polycrystalline $MgB_{2}$ samples with 
$RRR$ values between 2 and 8 (depending on the conditions of synthesis). The 
$T_{c}$ values as function of $RRR$ are shown in Fig. 2. Although the
variance in $T_{c}$ is less than $2\ K$ there is a clear tendency of $T_{c}$
increasing with $RRR$ in accordance with similar data for A15 thin film
superconductors.\cite{25,26} Extending the range of $T_{c}$ and $RRR$ by
including available data on $MgB_{2}$ films and neutron irradiated bulk
samples it was shown very recently that the Testardi correlation is
fulfilled over a wider range of $T_{c}$ and $RRR$ values\cite{2} and an
explanation was proposed in form of a defect-induced weak localization
correction to the electron-phonon coupling constant.\cite{27} Further
evidence for a high defect density should be found in structural parameters,
e.g. an increase of the lattice constant due to defects was observed in the
A15 compounds.\cite{26} It is also expected that the broader distribution of
(local) lattice parameters due to the presence of defects causes a
broadening of x- ray peaks. This effect can be extracted from x-ray spectra
and is quantitatively characterized by the microstrain, $\epsilon $. Fig. 3
shows both, the lattice parameters $a$, $c$ and $\epsilon $ as a function of 
$RRR$ (a more detailed discussion is given in ref. 28). Whereas $c$ shows
the expected increase with decreasing $RRR$ (i.e. increasing defect density)
a similar clear tendency is missing in the lattice parameter $a$. Obviously,
defects in $MgB_{2}$ mainly affect the inter-plane spacing. The microstrain
is strongly increasing because of the larger (local) distortion of the
lattice at the higher defect content (Fig. 3 B). The results of this Section
show that the majority of the $MgB_{2}$ bulk samples and thin films (with
low $RRR$) have a high defect concentration giving rise to the differences
in $T_{c}$, $RRR$, and structural parameters reported by various groups in
recent publications.

\subsection{Pressure Effects on $T_{c}$: Current Understanding and Unsolved
Problems}

Soon after the discovery of superconductivity in $MgB_{2}$ there was the
speculation\cite{6} that the application of pressure might enhance $T_{c}$
well above 40 $K$ similar to what has been found more than a decade ago in
high $T_{c}$ cuprate superconductors. Unfortunately, the first pressure
experiments clearly indicated a suppression of $T_{c}$ with hydrostatic
pressure.\cite{17,19} This result was confirmed by a number of subsequent
high-pressure investigations.\cite{29,30,31,32,33,34,35} Whereas all reports
agree about the negative sign of $dT_{c}/dp$, the differences in the value
are still a matter of discussion. $T_{c}$ decreases linearly with $p$ in the
low-pressure range ($p<2\ GPa$) with a coefficient varying between $-1\
K/GPa $ and $-2\ K/GPa$.\cite{17,29,30,31,32,33,34} Experiments on $MgB_{2}$
in an extremely non-hydrostatic environment yield an even lower value ($-0.3$
to $-0.8\ K/GPa$) for the pressure coefficient of $T_{c}$.\cite{19,35} At
higher pressure ($p>10\ GPa$) deviations from linearity become obvious.\cite
{32,33,34} The order of magnitude and the sign of $dT_{c}/dp$ are well
explained by the strong coupling BCS theory. Calculations of the pressure
effect on the band structure and the electron-phonon coupling found
reasonable agreement with the experimental data.\cite{36} Using equation (4)
the pressure coefficient is calculated as

\begin{equation}
\frac{d\ln T_{c}}{dp}=\frac{d\ln (\omega _{\ln })}{dp}+\frac{1.04\lambda
(1+0.38\mu ^{\ast })}{\left[ \lambda \left( 1-0.62\mu ^{\ast }\right) -\mu
^{\ast }\right] ^{2}}\left\{ \frac{d\ln N\left( E_{F}\right) }{dp}-\frac{%
d\ln \left\langle \omega ^{2}\right\rangle }{dp}\right\}  \eqnum{5}
\end{equation}

Thereby, any pressure dependence of $\mu ^{\ast }$ and $\left\langle
I^{2}\right\rangle $ was neglected. With reasonable values for $\mu ^{\ast
}\thickapprox 0.1$ and $\lambda \thickapprox 0.7$ Loa and Syassen\cite{36}
found good agreement of the calculated $dlnT_{c}/dp$ with the available
data. It is interesting to note that both terms in the wavy brackets of (5)
give a negative contribution ($N(E_{F})$ decreases and the average phonon
frequency increases with pressure) but the change of $\left\langle \omega
^{2}\right\rangle $ dominates the pressure effect.

There is an ongoing discussion about the very different values of $dT_{c}/dp$
reported so far. Tomita et al.\cite{30} speculated that the pressure
coefficient might be sensitive to the pressure medium used in the
experiments and to shear stress possibly introduced by less hydrostatic
pressure conditions resulting in a larger absolute value (closer to $2\ K/GPa
$) of $dT_{c}/dp$. Contrary to this assumption, the smallest values of $%
\left| dT_{c}/dp\right| $ have been obtained using the most non-hydrostatic
media, steatite.\cite{19,35} It should be noted that all pressure media
(except He below $0.5\ GPa$) freeze above $T_{c}$ of $MgB_{2}$. We have
measured different $MgB_{2}$ samples at hydrostatic (He gas pressure) and
quasi-hydrostatic (Fluorinert liquid) conditions and did not find a strong
sensitivity of $dT_{c}/dp$ to the pressure medium.\cite{17,31} In
particular, we found different pressure coefficients for different samples
under the same hydrostatic pressure conditions.\cite{31} No change of $%
dT_{c}/dp$ was observed in our He gas pressure experiments in passing
through the 0.5 GPa threshold above which the He freezes above $T_{c}$.
However, our experiments clearly show a correlation of the pressure
coefficient and the ambient pressure $T_{c}$ (Fig. 4). Samples with smaller $%
T_{c}$ tend to have a larger pressure coefficient. Additional data from
literature are also included in Fig. 4 (open symbols) and support the
correlation between $T_{c}$ and the pressure coefficient. Based on the
discussion of the previous Section the lower $T_{c}$ reflects a higher
degree of defects and distortions of the $MgB_{2}$ structure. Park et al. 
\cite{27} suggested that the defects mainly cause a reduction of the
electron-phonon coupling constant, $\lambda $, due to weak localization
effects. One may raise the question if the same reduction of $\lambda $ can
also explain the observed differences in the pressure coefficient. The
maximum change of $T_{c}$ of about $2\ K$ corresponds to an increase of $%
dT_{c}/dp$ by a factor of 2 (from $-1\ K/GPa$ to $-2\ K/GPa$, Fig. 4). Using
equation (4) with reasonable parameters, $\lambda =0.7...1$ and $\mu ^{\ast
}=0.1...0.13$, we estimate that $\lambda $ decreases by about $2$ to $3.5\ \%
$ for a $2\ K$ drop of $T_{c}$. Using equation (5) and the estimated values
for $dln\omega /dp\thickapprox 0.71\ \%/GPa$ and $dln\lambda /dp\thickapprox
-1.7\ \%/GPa$\cite{36} we calculate the change of $dT_{c}/dp$ if it is
solely due to the decrease of $\lambda $. We find that the expected change
of the pressure coefficient is negligibly small ($<1\ \%$) for the
reasonable values of $\lambda $ and $\mu ^{\ast }$ listed above. Therefore,
the large differences of $dT_{c}/dp$ observed for different $MgB_{2}$
samples and the correlation with $T_{c}$ (Fig. 4) cannot be explained by the
decrease of the ambient pressure $\lambda $ due to defect-induced weak
localization effects.\cite{27} Other mechanisms have to be considered to
understand the high-pressure data.

One possibility we may speculate about is an increase of defect
concentration by pressure. This would give rise to an additional suppression
of $\lambda $ and $T_{c}$ with pressure. The effect could be small for
samples with higher $T_{c}$ (or lower initial defect density) but larger for
''poor'' samples with a high defect concentration at ambient conditions. In
fact, recent experiments have shown that application of high pressure may
even irreversibly change the superconducting properties, particularly $T_{c}$%
.\cite{37} Careful structural characterization of $MgB_{2}$ at high pressure
should give additional insight into the role of defects and how they are
affected by external pressure.

\acknowledgments
This work was supported in part by NSF Grant No. DMR-9804325, MRSEC/NSF
Grant No. DMR- 9632667, the T. L. L. Temple Foundation, the John and Rebecca
Moores Endowment, the State of Texas through the Texas Center for
Superconductivity at the University of Houston, and at Lawrence Berkeley
Laboratory by the Director, Office of Energy Research, Office of Basic
Sciences, Division of Material Sciences of the U. S. Department of Energy
under Contract No. DE- AC0376SF00098.

%
%
\begin{figure}[tbp]
\caption{Seebeck coefficient of $Mg_{1-x}Al_xB_2$. The curves for different $%
x$ are offset by a constant and the zero values are indicated by short
horizontal lines}
\label{Fig. 1}
\end{figure}

\begin{figure}[tbp]
\caption{$T_c$ of various $MgB_2$ samples as function of the residual
resistance ratio. The line is a guide to the eye reflecting the Testardi
correlation over a wider range of $T_c$ and $RRR$.}
\label{Fig. 2}
\end{figure}

\begin{figure}[tbp]
\caption{Lattice constants (A) and micro strain (B) of $MgB_2$ as function
of the residual resistance ratio.}
\label{Fig. 3}
\end{figure}

\begin{figure}[tbp]
\caption{Pressure coefficient as function of $T_c$. Open circles are
literature data from ref. 29, 30, 32, 33.}
\label{Fig. 4}
\end{figure}

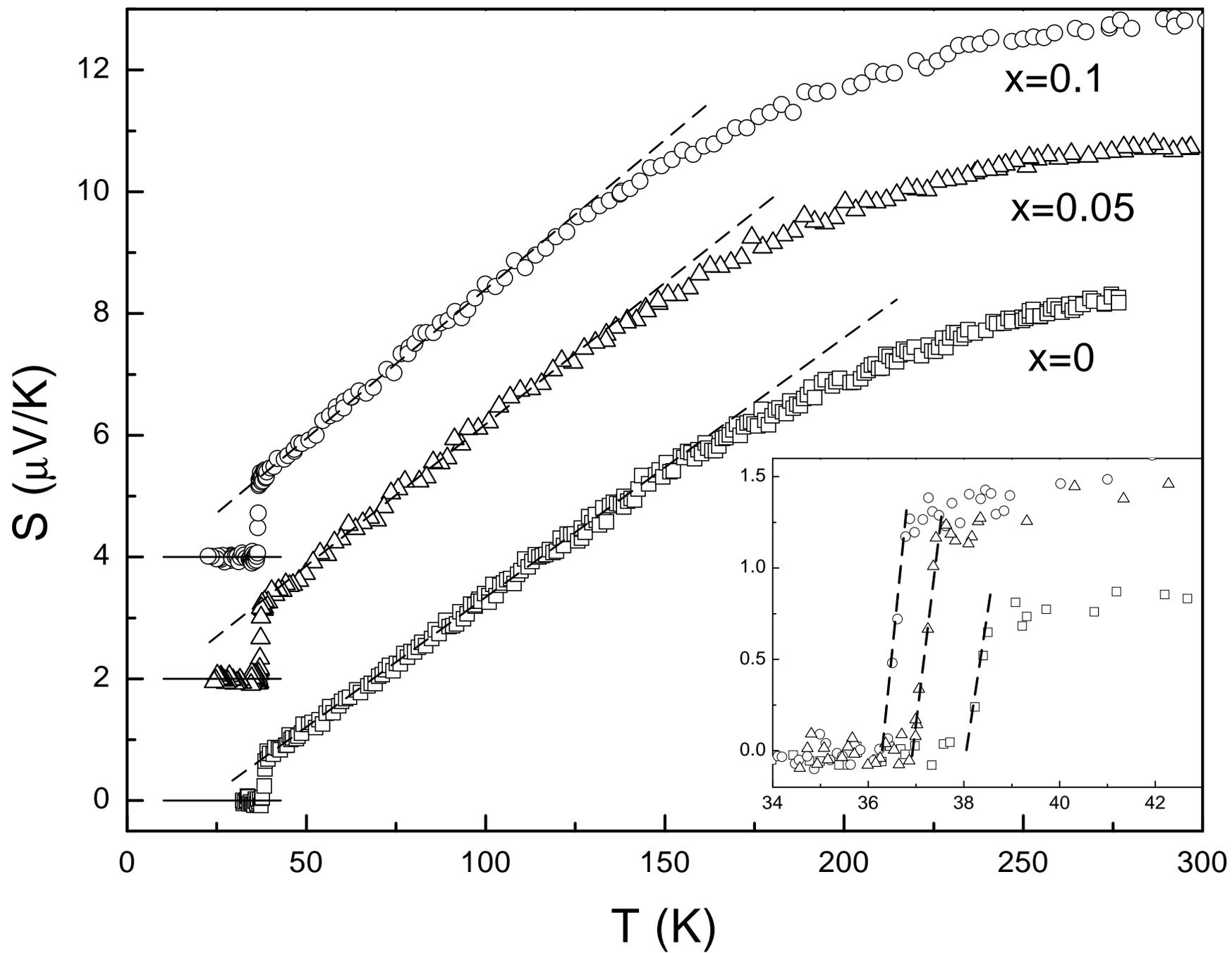
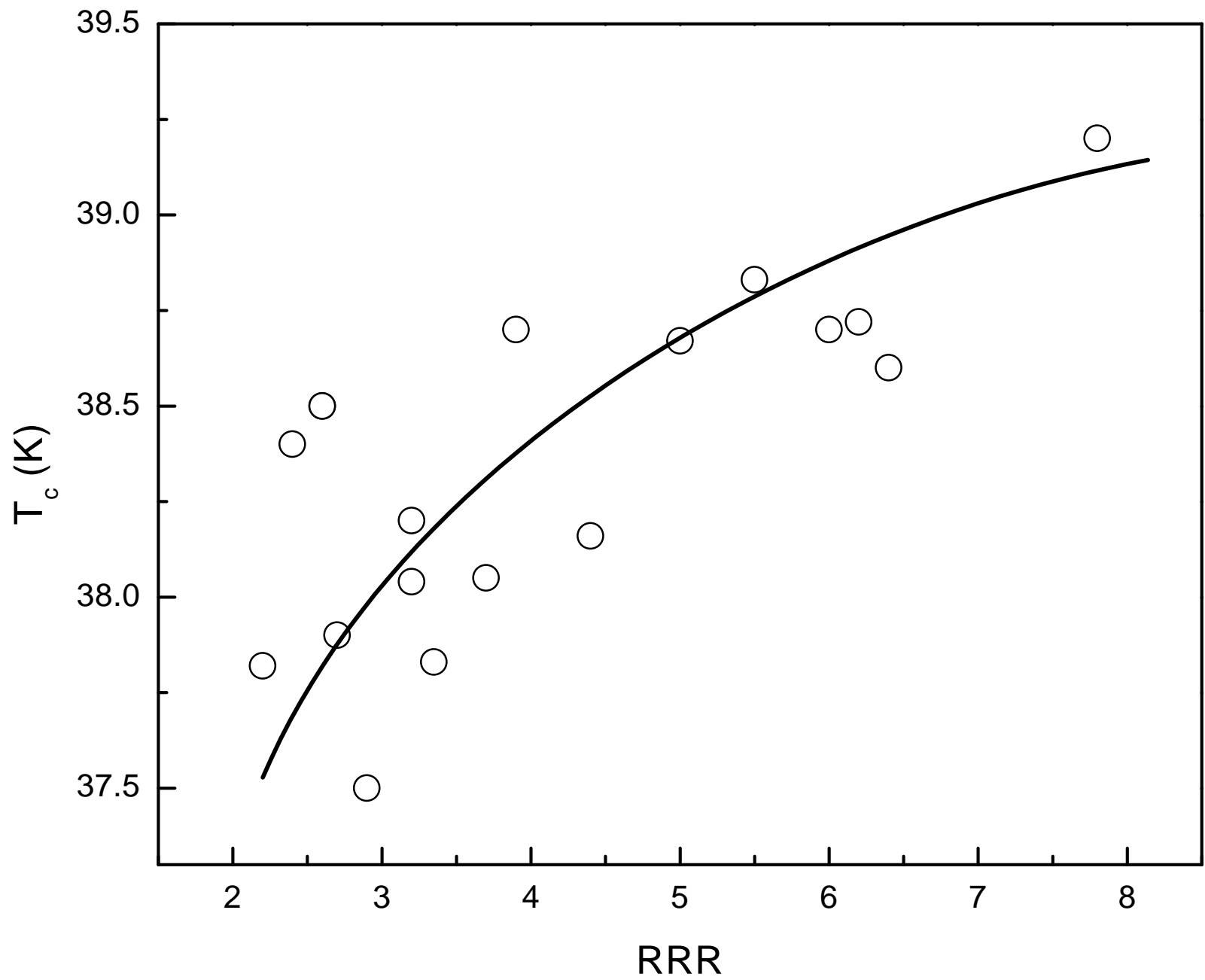
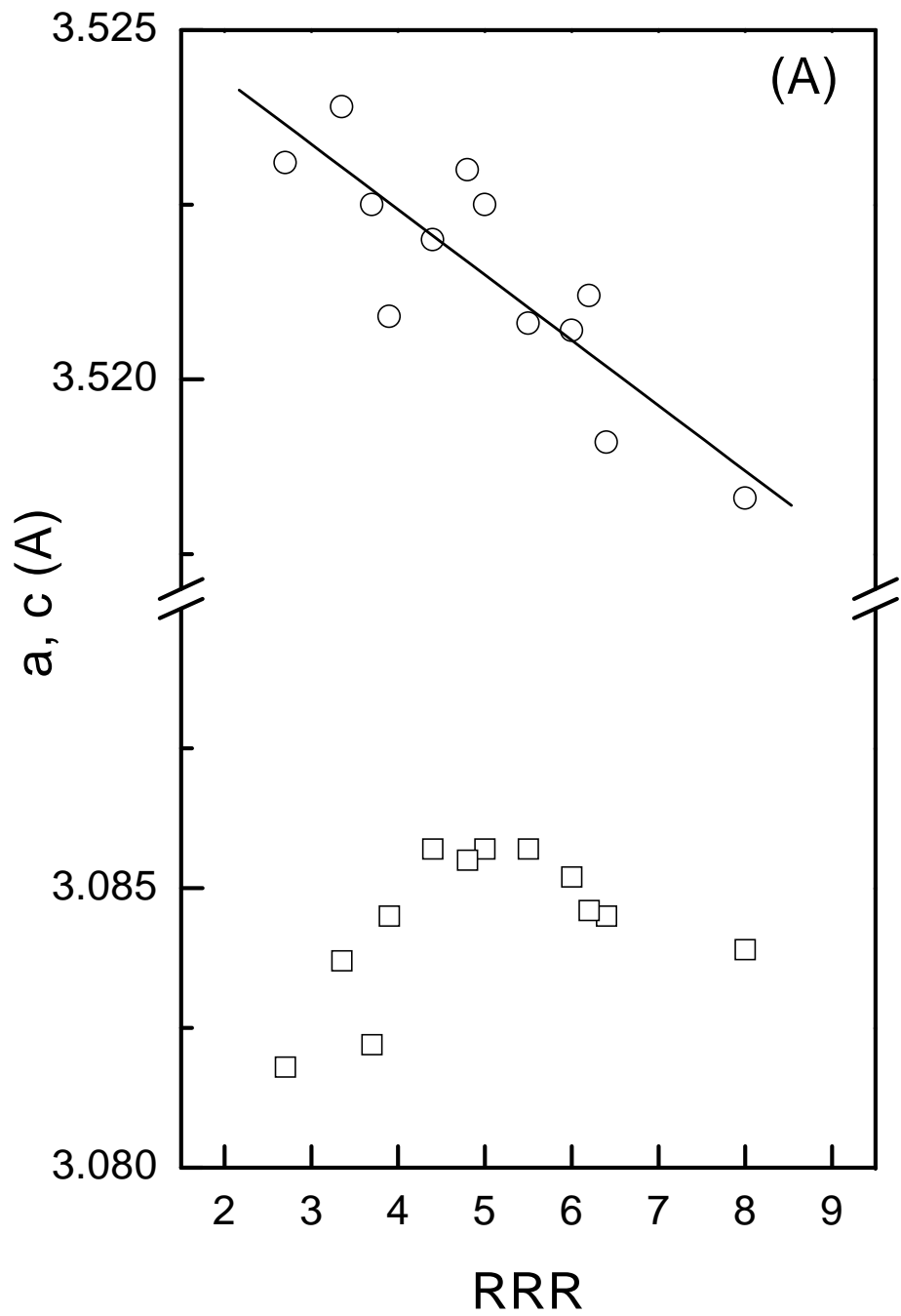
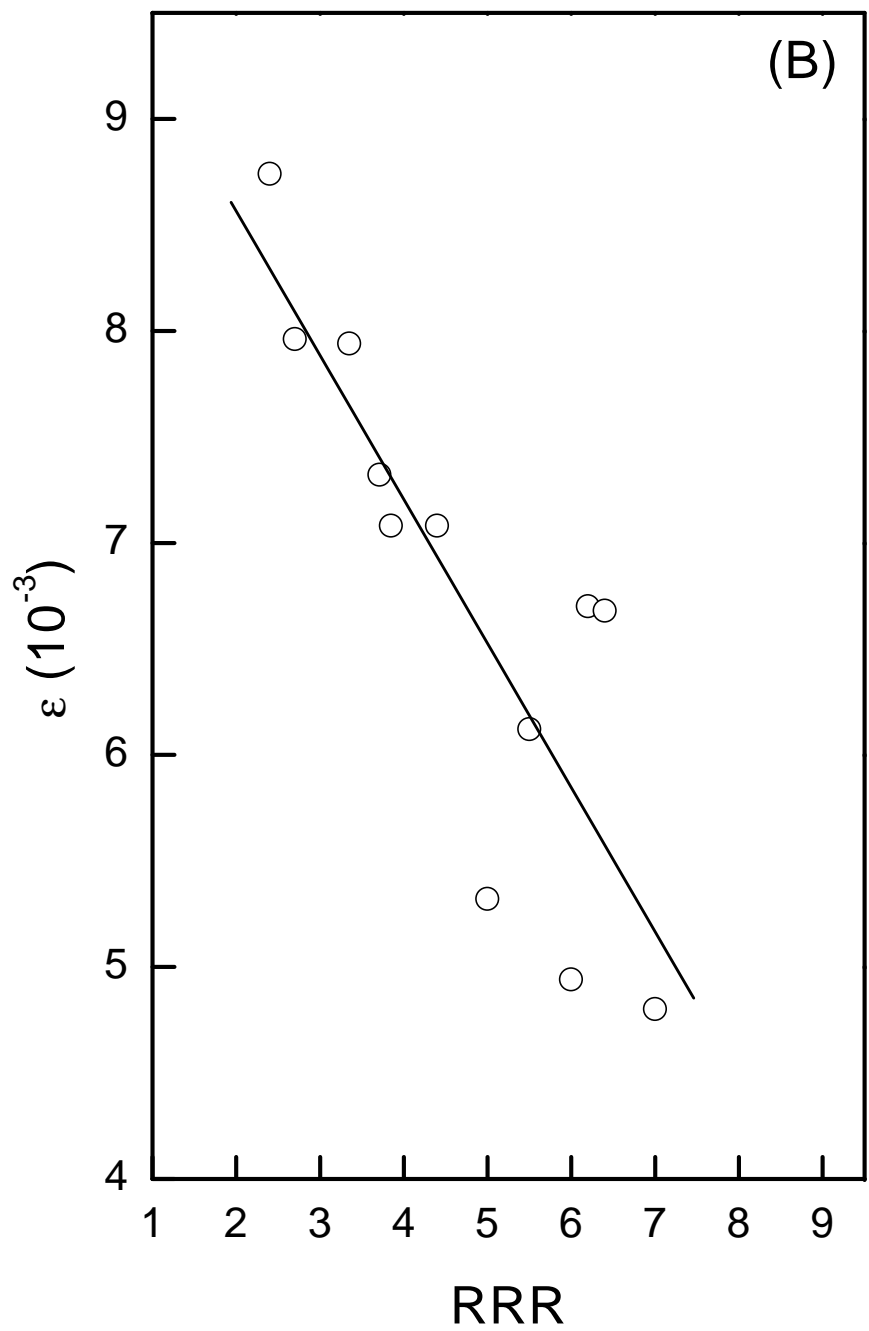
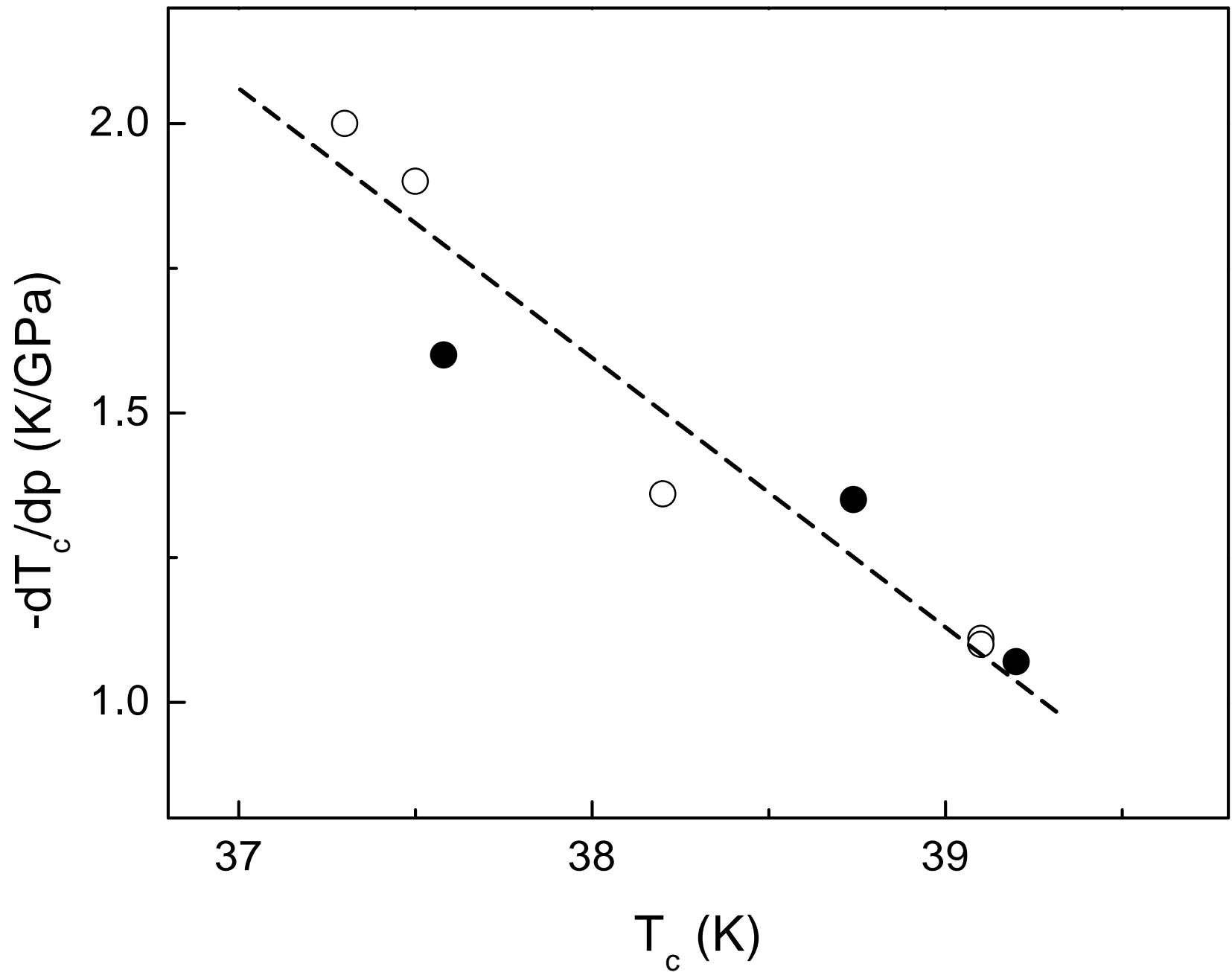
\end{document}